# Strong Coupling of the Mo–Mo Stretching Mode to a Locally Confined Stokes Raman Field in $Mo_2$ Molecular Resonators

Miao Meng, Ying Ning Tan, Guang Yuan Zhu, Chun Y. Liu*

**Affiliations:**

Department of Chemistry, College of Chemistry and Materials Science, Jinan University, 601 Huang–Pu Avenue West, Guangzhou 510632, China
Correspondence to: tcyliu@jnu.edu.cn

**Abstract**

Confining optical fields to molecular dimensions is a central objective in nanophotonics and molecular quantum optics. Here, we report strong coupling of the Mo–Mo stretching vibration to a locally confined Raman scattering field in quadruply bonded dimolybdenum complexes. Under ambient, cavity-free conditions, the dimolybdenum formamidinate complexes $Mo_2(DAniF)_4$ and $Mo_2(DTolF)_4$ exhibit Rabi-type splitting, Mollow-type sidebands, and higher-order Raman features centered near the Mo–Mo stretching frequency of 400 cm$^{-1}$, indicating formation of dressed vibration–field states. The sideband displacements from the resonance follow the photon-number-dependent relation $\Omega_n = \Omega_v\sqrt{n}$, consistent with Jaynes–Cummings-type coupling, while strongly displaced Raman features are assigned to leapfrog transitions within the same dressed-state ladder. In contrast, the less polarizable $Mo_2(O_2CCH_3)_4$ complex exhibits essentially a single Mo–Mo stretching band. Reanalysis of reported resonance-Raman spectra of an alkynyl $Mo_2$ complex further supports this coupling framework. These results suggest that the $Mo_2$ unit simultaneously serves as the Raman-active oscillator and the molecular resonator that supports, confines, and enhances the locally generated scattering field, providing spectroscopic evidence for vibration–field coupling and optical-field confinement within a chemically defined metal–metal bond.

## Introduction

Strong vibrational coupling provides a direct route for integrating molecular nuclear motion with electromagnetic (EM) fields, producing hybrid vibro-polaritonic states whose properties differ from those of the uncoupled vibrational and photonic modes. [1,2,3,4,5,] Because molecular vibrations are intrinsically associated with specific bonds and structural coordinates, vibration–field coupling offers a particularly direct means of modifying molecular energy flow, vibrational dynamics, and potentially chemical reactivity at the level of selected chemical bonds. Conventionally, such coupling is achieved by placing molecular ensembles inside optical or infrared cavities whose photonic resonances are tuned to characteristic vibrational transitions.[1,2,3] The resulting hybridization introduces a photonic degree of freedom into molecular nuclear motion and can modify vibrational relaxation, molecular structure, and chemical dynamics.[3,6,7,8,9,10,11] These developments have established vibration–field coupling as an important interface between molecular spectroscopy, quantum optics, and polaritonic chemistry.

A central challenge is to reduce the spatial scale of the EM mode to sustain sufficiently strong interaction with a molecular vibration. Plasmonic nanocavities concentrate optical fields into deeply subwavelength volumes and have enabled strong single-molecule coupling and greatly enhanced Raman responses.[12,13,14,15,16,17,18] Surface- and tip-enhanced Raman spectroscopy further demonstrate that highly localized near fields can amplify the optical response of individual molecular vibrations, while atomic protrusions in metallic junctions form picocavities whose effective mode volumes approach molecular dimensions.[13,15,16,17,18,19,20,21] Quantum molecular optomechanics provides a complementary description of this light-matter coupling regime, treating Raman scattering as an energy-exchange process between molecular vibrations and confined EM fields.[13,22] In particular, quantum treatments of plasmon-enhanced Raman scattering explicitly describe nonresonant Stokes scattering as a two-photon transition between molecular vibrational states, while experiments show that Stokes process can “pump” the $\nu_0 \rightarrow \nu_1$ transition and populate a continuum of the vibrational excited states.[18, 23] These advances progressively connect Raman spectroscopy with nanoscale light–matter coupling, but the required optical confinement has so far been supplied predominantly by externally fabricated metallic or dielectric structures. This raises a more fundamental question: can a chemically defined molecular structure itself generate, confine, and enhance a local scattering field sufficiently to couple coherently to one of its own characteristic vibrations?

Our previous studies of dinuclear $Ni_2$ and $Mo_2$ complexes suggest such an intrinsic molecular-resonator architecture to be the best candidate for implement of this mission.[24,25,26,27] In these systems, optical excitation of the dimetal unit produces resonance-fluorescence features including Rabi doublets and Mollow-type triplets, which were interpreted within a Jaynes–Cummings (JC) molecular framework in which a discrete molecular excitation interacts with a locally generated and quantized scattering field.[25,26,27] The recently characterized $Ni_2$ system provides a particularly well-resolved example, exhibiting strong light–matter coupling under weak excitation in free space and supporting the view that two closely spaced metal centers can function as an integrated emitter–resonator quantum system.[25] The ångström-scale intermetallic geometry defines an exceptionally small effective field-confinement region, conceptually extending the progression from plasmonic nanocavities and picocavities toward a molecular optical resonator. An important unresolved question, however, is whether this resonator functionality is restricted to electronic excitation–field coupling or can also act on a structurally defined nuclear coordinate. A characteristic molecular vibration localized directly on the dimetal unit therefore provides a stringent internal probe of the proposed field-confinement mechanism.

Quadruply bonded $Mo_2$ complexes are particularly suitable for this purpose. Their short Mo–Mo bonds possess a strong and structurally well-defined, Raman-active stretching vibration, typically near 400 $cm^{-1}$, that has long served as a characteristic spectroscopic signature of the metal–metal quadruple bond.[28,29,30,31,32,33] Unlike an electronically assigned optical transition, the Mo–Mo stretching mode directly probes the same intermetallic coordinate that is proposed to define the molecular resonator for a terahertz field. Moreover, its Raman response is strongly sensitive to the electronic structure and polarizability of the $Mo_2$ core. Supporting ligands can therefore provide a chemical means of tuning the Raman activity and local-field response while largely preserving the underlying Mo–Mo structural motif. This combination of a bond-specific Raman oscillator, an ångström-scale dimetal architecture, and ligand-tunable polarizability makes $Mo_2$ complexes an desirable platform for testing whether a locally generated Raman scattering field can become sufficiently confined and enhanced to enter the coherent vibration–field coupling regime.

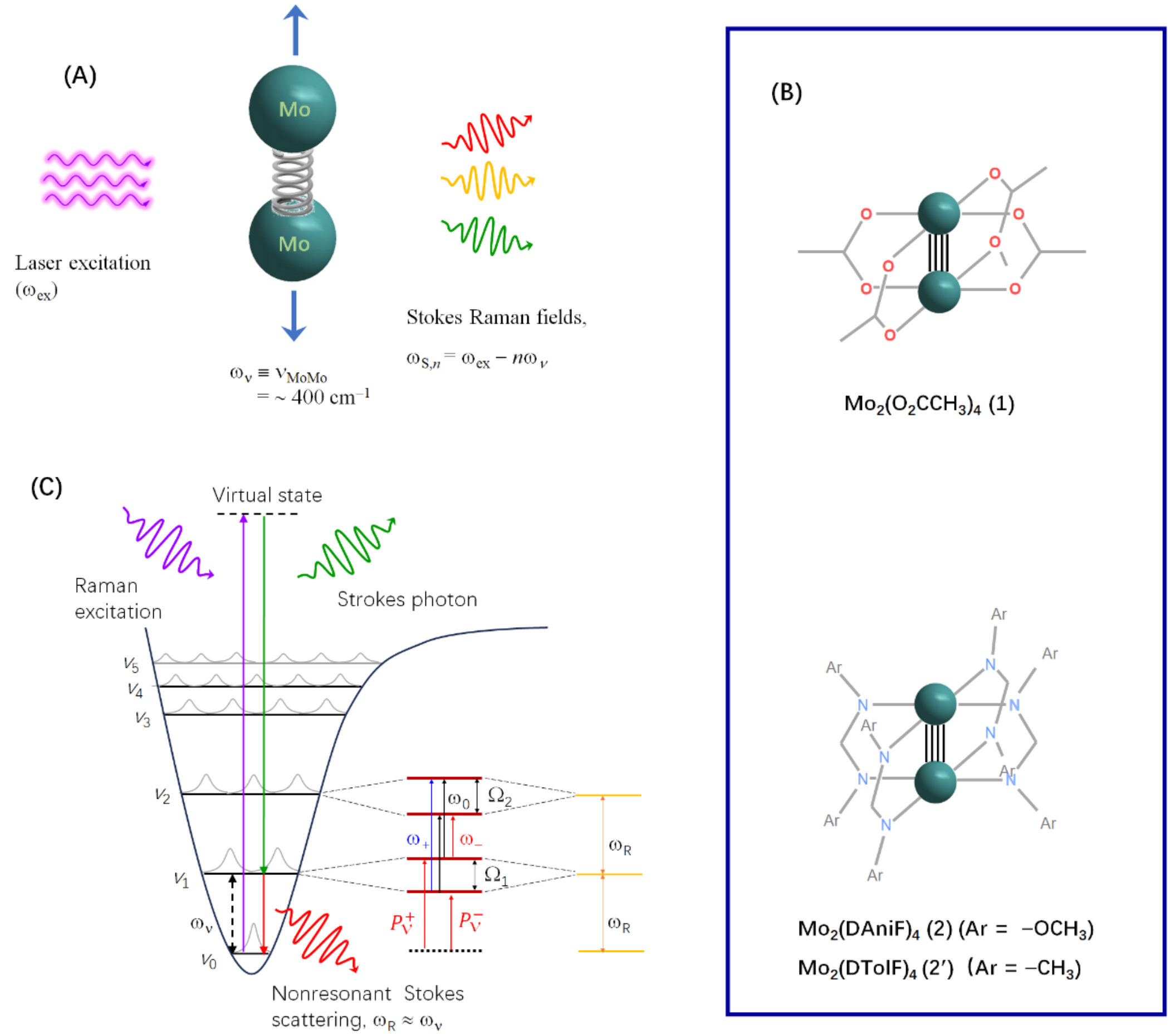

**Figure 1. Molecular system and proposed vibration–field coupling framework.** (A) Schematic illustration of the Raman energy-transfer channel generated upon optical excitation of the quadruply bonded $Mo_2$ unit. Laser excitation at frequency $\omega_{ex}$ produces nonresonant Stokes Raman transitions with Raman energy shifts $n\hbar\omega_v$, where $n$ denotes the vibrational quantum index and $\omega_v$ is the Mo–Mo stretching frequency near 400 cm$^{-1}$. In the proposed $Mo_2$ molecular-resonator picture, the associated locally confined Raman scattering

response is represented by an effective quantized field mode of fundamental frequency ($\omega_R$ $\approx \omega_v$, with Fock-state energies $E_{S,n} = n\hbar\omega_R$. The $Mo_2$ unit is proposed to confine and enhance this local Raman field at the intermetallic coordinate. (B) Molecular structures of $Mo_2(O_2CCH_3)_4$ (**1**), $Mo_2$(DAniF)$_4$ (**2**), and $Mo_2$(DTolF)$_4$ (**2′**). (C) Schematic formation of the vibro-polaritonic manifolds through resonant coupling between the Mo–Mo stretching coordinate and the effective local Raman field mode $\omega_R \approx \omega_v$. The lowest coupled manifold gives the lower and upper Rabi-type branches, $P_v^-$ and $P_v^+$, whereas transitions involving populated higher field manifolds give rise to a central component and Mollow-type sidebands.

Here, we investigate the Raman spectra of $Mo_2(O_2CCH_3)_4$ (**1**), $Mo_2$(DAniF)$_4$ (**2**), and $Mo_2$(DTolF)$_4$ (**2′**), which contain closely related quadruply bonded $Mo_2$ cores but substantially different ligand environments. Whereas **1** exhibits essentially the conventional single Mo–Mo stretching band, the more polarizable formamidinate complexes **2** and **2′** display Rabi-type splitting, Mollow-type sidebands, and additional higher-order Raman features centered on the Mo–Mo stretching frequency. We interpret these observations using a Stokes Raman scattering coupling framework in which nonresonant Stokes Raman transitions establish the vibrational energy scale, while the associated locally confined scattering response of the $Mo_2$ resonator is represented by an effective field mode $\omega_R \approx \omega_v$. Resonant interaction of this effective Raman-field mode with the Mo–Mo stretching transition produces dressed vibro-polaritonic manifolds whose coupling displacements follow the characteristic square-root JC scaling, $\Omega_n = \Omega_v\sqrt{n}$. Higher-order Raman features are attributed to the leapfrog-type transitions through virtual intermediate states within the same dressed-state ladder. By integrating the Raman-active oscillator, the source and confinement of the local scattering field, and the molecular structure within a single crystallographically defined metal–metal bond, the present system extends vibration–field coupling from externally constructed cavities toward the structural limit of an intrinsic molecular optical resonator.

## Results

**Mollow-type and Rabi-type sidebands of the Mo–Mo vibrational mode observed for the dimolybdenum formamidinates 2 and 2′.** Illumination of a single crystal of $Mo_2(O_2CCH_3)_4$ (**1**) with a 532 nm laser beam produces a single intense Raman band at 410 cm$^{-1}$, assigned to the Mo–Mo stretching mode, $\omega_v$, as shown in Figures 2A and Figure S1. The surrounding spectral region from approximately 350 to 650 cm$^{-1}$ otherwise shows a nearly featureless baseline. The Raman scattering of the Mo–Mo stretching vibration in quadruply bonded dimolybdenum complexes has been extensively studied; $Mo_2(O_2CCH_3)_4$ is the prototype in this regard. In different measurements of **1**, a single-band feature has been commonly observed, although the mode frequency may shift within the range of 400 – 410 cm$^{-1}$.[28,30,33] For other $Mo_2$ complexes, their Mo-Mo stretching frequencies vary moderately with the supporting ligands and donor atoms, and the Raman spectra may exhibit additional scattering near the vibrational fundamental. Neighboring bands in the vicinity of $\omega_v$ have previously been assigned to overtones or mixed skeletal vibrations.[32,34] The spectrum of **1** therefore provides a useful reference for the conventional Raman response of an essentially isolated Mo–Mo stretching mode.

In marked contrast, excitation of powdered $Mo_2(DAniF)_4$ (**2**) at 532 nm produces a structured Raman envelope centered near 405 cm$^{-1}$ (Figures 2B and S2). In addition to the central Mo–Mo stretching band, two prominent features occur at approximately 377 and 432 cm$^{-1}$. These bands are nearly symmetrically displaced from the central frequency by about 27 cm$^{-1}$, giving a triplet-like spectral pattern. The symmetric placement of the sidebands is difficult to reconcile with a simple set of unrelated normal modes, because independent molecular vibrations would not generally be expected to occur at equal positive and negative displacements from the Mo–Mo stretching frequency. Instead, the pattern resembles a central resonance accompanied by a pair of coupling-induced sidebands, e.g., Rabi or Mollow sidebands.[1,2,35,36] Several broader and weaker bands are also observed farther from $\omega_v$, indicating that the spectral structure is not restricted to this innermost triplet.

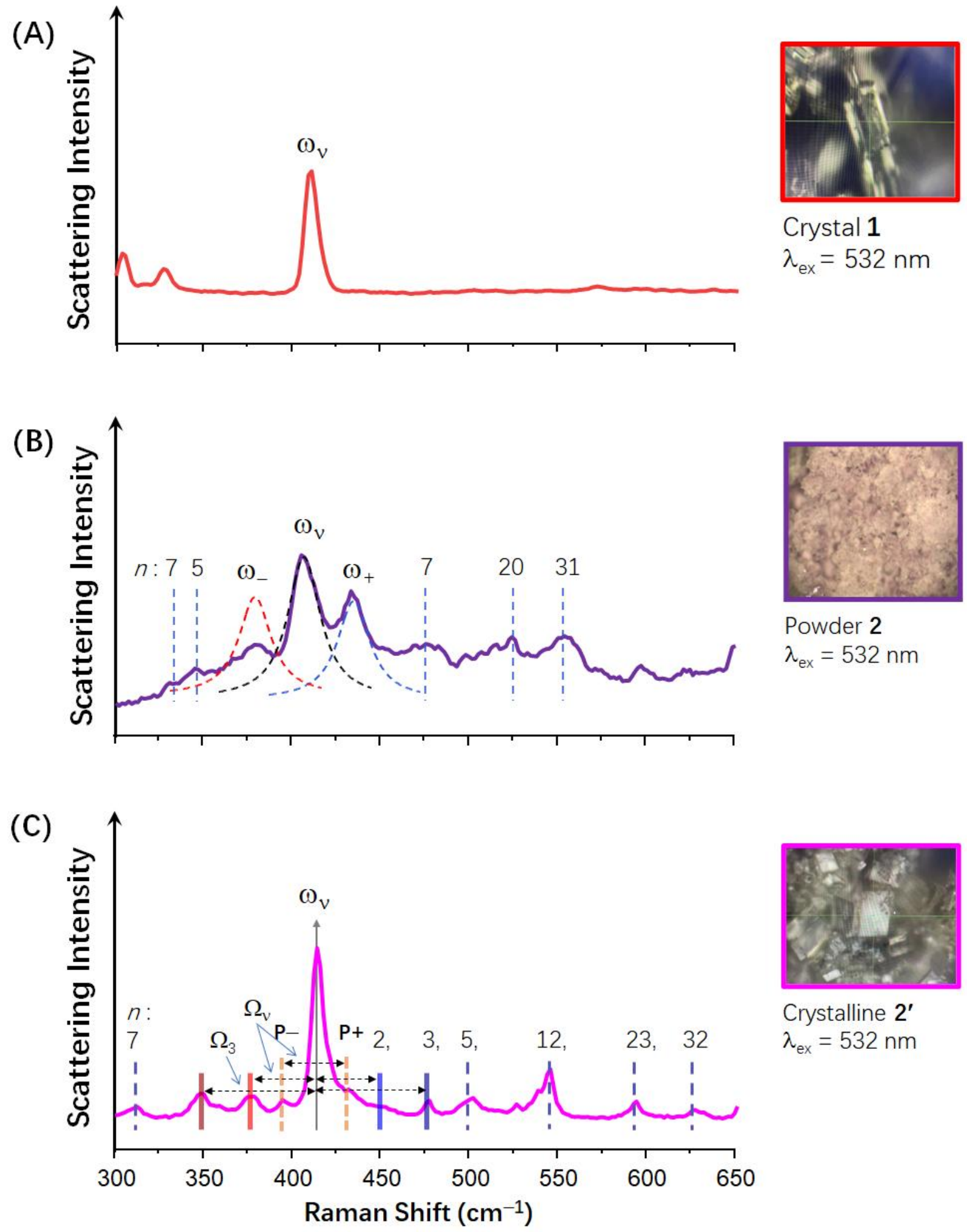

**Figure 2. Raman spectra of $Mo_2(O_2CCH_3)_4$ (1), $Mo_2(DAniF)_4$ (2), and $Mo_2(DTolF)_4$ (2′) obtained with 532 nm excitation.** (A) Raman spectrum of a single crystal of **1**, showing an intense isolated Mo–Mo stretching band, $\omega_\nu$, at 410 cm⁻¹. (B) Raman spectrum of powdered **2**, showing the central Mo–Mo stretching band near 405 cm⁻¹, an approximately symmetric inner sideband pair at 377 and 432 cm⁻¹, and additional higher-frequency and lower-frequency features. Dashed curves show the central and sidebands of the Mollow-type triplet in the spectrum; vertical dashed lines mark the effective field-manifold indices ($n$) that correspond to the sideband displacements ($\Omega_n$) according to $\Omega_n = \Omega_1\sqrt{n}$. (C) Raman spectrum of crystalline **2′**, showing the central Mo–Mo stretching band near 414 cm⁻¹, the inner Rabi-type branches $P_\nu^-$ and $P_\nu^+$, a Mollow-type sideband pair, and additional higher-order Raman features.

An even more resolved pattern is observed for crystalline $Mo_2(DTolF)_4$ (**2′**). Excitation at 532 nm gives an intense central band at approximately 414 cm⁻¹, together with several pairs of features distributed on both sides of the Mo–Mo stretching frequency (Figures 2C and S3). The innermost pair occurs at approximately 394 and 432 $cm^{-1}$ and is displaced from the central band by about ± 19 $cm^{-1}$. A second pair appears at approximately 375 and 451 $cm^{-1}$, corresponding to displacements of about ± 38 $cm^{-1}$. Thus, the displacement of the second pair is approximately twice that of the first. The innermost components may therefore be identified phenomenologically as a Rabi-type doublet,[1,2,3,4] $P_v^-$ and $P_v^+$, separated from the bare vibrational resonance by $\pm\ \frac{\Omega_v}{2} \approx \pm 19\ cm^{-1}$, whereas the second pair forms a Mollow-type sideband pair at $\omega_v \pm \Omega_v$. Additional features at larger Raman shifts indicate the presence of still higher-order spectral components.

The contrast among **1**, **2**, and **2′** is notable because all three complexes contain closely related quadruply bonded $Mo_2$ cores, while their ligand environments differ substantially. Complex **1** contains harder carboxylate oxygen donors, whereas **2** and **2′** contain softer, more electron-donating and resonance-delocalized formamidinate ligands. The more electron-donating and resonance-delocalized formamidinate ligands increase the electronic deformability of the $Mo_2$ core and enhance the derivative of its polarizability with respect to the Mo–Mo stretching coordinate. This increase in Raman activity accounts for the much greater intensity and spectral visibility of the Mo–Mo-centered features of the formamidinate complexes. It may also strengthen the coherent coupling of the vibrational coordinate with the local scattering field, although the spectra considered here do not by themselves establish whether the increased polarizability affects only the Raman intensity or also the coupling strength.

These Raman features cannot be adequately described as a single bare vibration accompanied by only conventional overtones or coupling between normal modes. At the same time, the sidebands exhibit unequal intensities, incomplete symmetric partners, and substantial broadening, unlike the idealized Rabi doublets commonly observed for vibrations coupled to a single externally defined cavity mode.[1,2,3,4,5] For both formamidinates **2** and **2′**, the coupling behavior observed in free space at room temperature without an external optical cavity raises questions such as: what optical field is resonant with the Mo–Mo vibration under ordinary Raman excitation? how can that field become to be sufficiently intense to drive the strong vibro-photon coupling and how should the splitting of the Mo-Mo stretching mode and the formation of more distant sidebands be interpreted? To address these issues and interpret the unusual Raman spectral profiles, a Stokes Raman scattering coupling mechanism is proposed in the following subsection.

**Resonant coupling of the Mo–Mo stretching mode to the Stokes Raman scattering field.** Conventional strong vibrational coupling is generally achieved by placing a molecular vibration within an external optical or infrared cavity whose photonic mode is tuned into resonance with the vibrational transition.[1,2,3,4, 5, 6,9,13,14,16] For the present light-hybrid $Mo_2$ system, we propose a Stokes Raman scattering coupling mechanism to account for the coherent vibration–field interaction and the Raman features associated with the Mo–Mo stretching mode. In the two-photon Raman process, nonresonant Stokes scattering occurs between two vibrational states of the $Mo_2$ molecule, ($\nu_0 \rightarrow \nu_1$),[18,23] with an energy difference ($\hbar\omega_v$). This scattering response of the molecule defines an effective quantized electromagnetic mode of fundamental frequency ($\omega_R \approx \omega_v$) (Figure 1B), which is locally confined by the $Mo_2$ resonator. [26,27] Under the resonance condition, the Fock-state energies $E_{R,n} = n\hbar\omega_R$ correspond to the quantized vibrational energy transferred to the molecule ($n\hbar\omega_v$). The vibrational transition $\omega_v$ is resonantly coupled to Raman-excitation "pumping" filed, [18,23] forming a ladder of the dressed vibro-polaritonic states (Figure 3). Thus, the Raman/vibrational quantum index ($n$) maps directly onto the photon-occupation index of the effective Raman-field mode. A closely related framework is provided by quantum molecular optomechanics in plasmonic nanocavities, in which Raman scattering is described as energy exchange between molecular vibrations and strongly confined EM fields.[13,18,22] Distinct from conventional vibration–cavity coupling, where the molecular vibration interacts with an externally defined photonic mode and the resulting vibro-polaritonic states are populated by an applied driving field, the effective Raman field in the present $Mo_2$ system is proposed to arise locally from the scattering process and to be enhanced by the molecular resonator itself. Raman excitation at $\omega_{ex}$ populates the dressed vibro-polaritonic manifolds through two-photon Raman transitions, with the detected Stokes-photon frequency satisfying $\omega_{S,n} = \omega_{ex} - n\omega_R$. Spontaneous Raman scattering from the populated dressed states then produces the observed Rabi-type and Mollow-type Raman features, as illustrated in Figure 3.

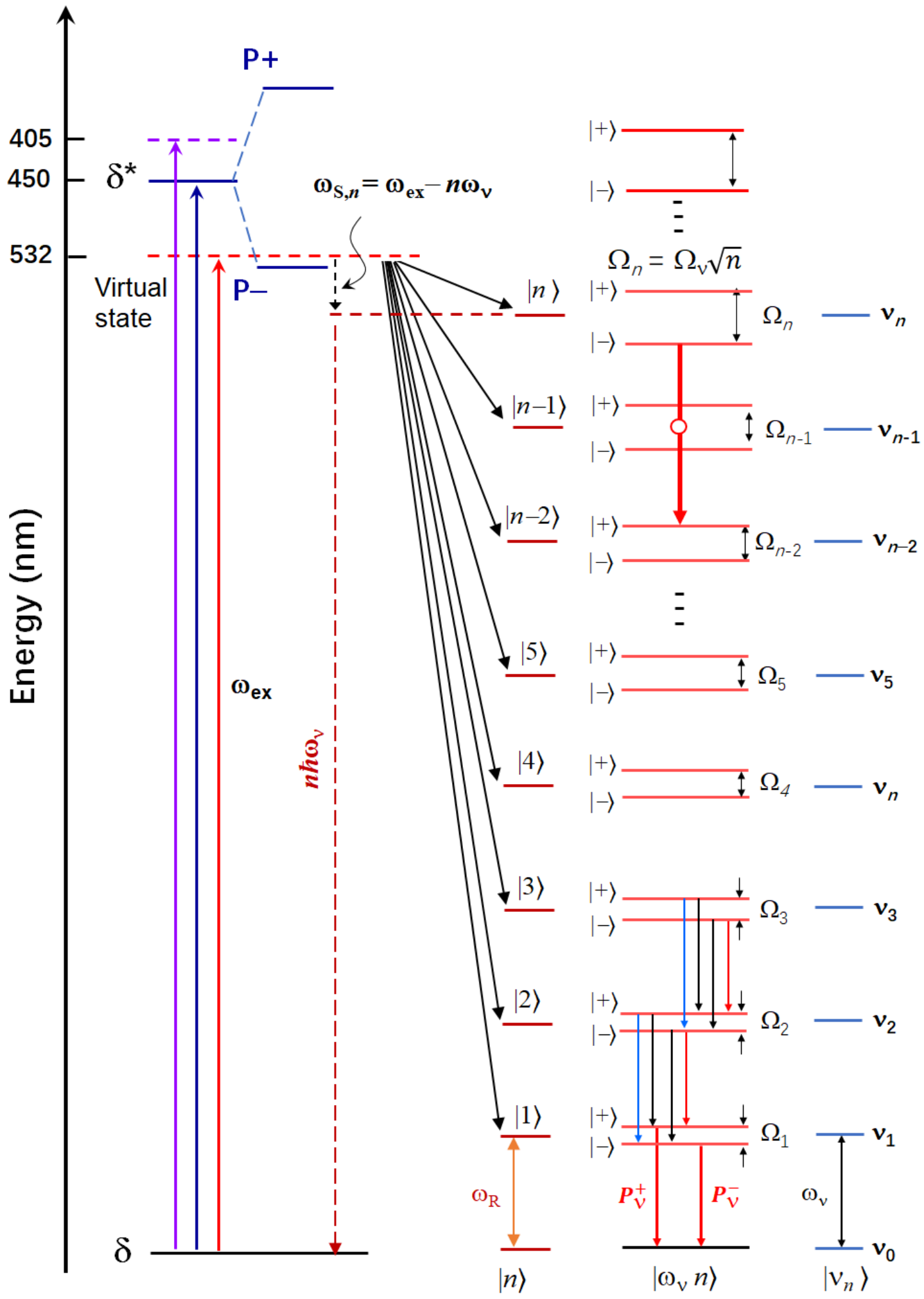

**Figure 3. Effective dressed-state ladder for coupling of the Mo–Mo stretching vibration to the local nonresonant Stokes Raman field in complexes 2 and 2′.** The Mo–Mo vibrational transition $\omega_v$ couples resonantly to the effective local Raman mode $\omega_R$, producing dressed vibration–field manifolds whose coupling displacements follow $\Omega_n = \Omega_v\sqrt{n}$, with $n$ = 1, 2, 3,... The dressed manifolds are populated by the Stokes Raman scattering at $\omega_{S,n} = \omega_{ex} - n\omega_v$. Spontaneous scattering from produces the observed Rabi-type and Mollow-type Raman features. Under 405 nm excitation, coupling to the δ→δ* electronic transition can additionally populate the upper electronic exciton-polaritonic state P+, facilitating access to the high-

energy vibro-polaritonic manifolds. The thick red arrow interrupted by an open circle indicates the $|-\rangle \rightarrow |\pm\rangle \rightarrow |+\rangle$ leapfrog-type transition via a virtual intermediate state (circle marked).

Therefore, the coupled system can be described phenomenologically by an effective Jaynes–Cummings-type interaction [37] between the Mo–Mo vibrational transition and the quantized local Raman-field mode at the single-molecule level. Within the rotating-wave approximation, the effective Hamiltonian may be written as[37,38,39]

$$H = \hbar\omega_R a^\dagger a + \frac{1}{2}\hbar\omega_v\sigma_z + \hbar g(a^\dagger\sigma_- + a\sigma_+)$$

where $a^\dagger$ and $a$ create and annihilate photons in the effective local Raman-field mode, $\sigma_+$ and $\sigma_-$ describe excitation and de-excitation of the effective Mo–Mo vibrational transition, $\sigma_z$ is the corresponding inversion operator, and g is the single-quantum vibration–field coupling strength. At resonance, $\omega_R \approx \omega_v$, the uncoupled vibration and field states mix to form dressed vibro-polaritonic manifolds. Accordingly, the effective coupling displacement associated with manifold $n$ is expressed as

$$\Omega_n = \Omega_v\sqrt{n} \quad (n = 0, 1, 2, 3\ldots),$$

where $\Omega_v$ is the fundamental coupling scale determined for the given $Mo_2$ system. This square-root dependence produces a nonlinear ladder of dressed vibration–field states, as illustrated in Figure 3, analogous to coupling a two-level excitation in the cavity quantum electrodynamics (QED).[38,39] At the lowest populated manifold ($n$ = 0), the coupling generates the lower and upper Rabi-type branches positioned symmetrically about the Mo–Mo vibrational resonance, i.e., $P_v^{\pm} = \omega_v \pm g$. Transitions involving higher dressed manifolds may produce a central component near $\omega_v$, accompanied by a series of $n$-dependent sidebands at $\omega_{SB,n} = \omega_v \pm \Omega_n$. These three-component structures are referred to here as Mollow-type triplets[35] because they contain a central Raman feature and coupling-related blue and red sidebands. Different from the vibration-cavity coupling, the experimental spectra frequently show incomplete triplets and asymmetric intensity distributions because of unequal manifold populations, competing relaxation pathways, spectral overlap, and structural inhomogeneity.

Experimentally, the displacement of a sideband from the central Mo–Mo stretching frequency gives $\Omega_n(\text{exp}) = |\omega_{SB,n}(\text{exp}) - \omega_v|$. The corresponding effective photon-number index ($n$) is then be estimated from $n = (\Omega_n/\Omega_v)^2$. The integer $n$ therefore labels the photon occupation of the effective local Raman-field mode within the coupling model. For the assignments in this and the related $Mo_2$ systems, the integer $n$ was selected such that the calculated and observed Raman positions differ by no more than 1.0 cm$^{-1}$: $|\omega_{SB,n} - [\omega_v \pm \Omega_n(\text{cal})]| \leq 1.0$ cm$^{-1}$, where $\Omega_n(\text{cal}) = \Omega_v\sqrt{n}$ and $n$ is determined from $\Omega_n(\text{exp})$. The observed and calculated Raman shifts and their residuals are summarized in Tables S1 and S2.[40]

Within this framework, the Raman features between 300 and 600 cm$^{-1}$ for powdered complex **2** can be assigned to transitions involving vibro-polaritonic manifolds with defined $n$ indices (Figure 2B and Table S1). The highest-energy band at 555 cm$^{-1}$ is displaced from $\omega_v$ by 150 cm$^{-1}$, giving $n = 31$. Additional features indicated by the dashed lines in Figure 2B are assigned to Mollow-type sidebands associated with effective photon-number manifolds $n$ = 5, 7, and 20 (Table S1). The breadth and asymmetry of these bands suggest that several closely spaced dressed-state transitions may contribute and that their populations are not uniform across the powdered sample. The high-order sidebands observed for complex **2′** (Figure 2C) provide further support for this coupling mechanism. In addition to the primary low-order structure, complex **2′** exhibits a pair of sidebands at 348 and 477 cm$^{-1}$, displaced from the central Mo–Mo resonance by approximately ± 65 cm$^{-1}$. Together with the central Raman feature at 413 cm$^{-1}$, these two bands form a higher-order Mollow-type triplet with $n$ = 3, in close agreement with the calculated values $\omega_{SB,n}$ (Table S1). Additional blue sidebands of **2′** are observed at approximately 500, 511, 545, 595, and 628 cm$^{-1}$. Their displacements from $\omega_v$ are reproduced by the same square-root scaling with $\Omega_v = 38$ cm$^{-1}$ and photon-number indices $n$ = 5, 12, 23 and 32, respectively. A lower-energy feature near 313 cm$^{-1}$ is assigned as the red sideband associated with $n$ = 7 (Figure 2C and Table S1). The highest assigned manifold, $n$ = 32, corresponds to $\Omega_{32} = \Omega_v\sqrt{32} \approx 215$ cm$^{-1}$.

The observation of such high-order features indicates that Raman excitation can populate multiple photon-number manifolds of the effective local Stokes scattering field.[23] Coupling of these manifolds to the Mo–Mo vibrational transition produces the experimentally resolved Jaynes–Cummings-type ladder of dressed vibro-polaritonic states.[37,38,39] The relative intensities of the central Mo–Mo band and its sidebands differ between complexes **2** and **2′**. The weaker sideband intensities of **2′** may be associated with the lower electron-donating and resonance-polarizing ability of the methyl substituent in DTolF (Figure 1B). By contrast, the enhanced polarizability for **2** with a methoxyl group on DAniF may increase both the Raman scattering efficiency and the population of the coupled vibration–field manifolds.

The proposed mechanism accounts for the simultaneous occurrence of an inner Rabi-type splitting, Mollow-type sidebands, and more distant higher-order Raman features centered on a common Mo–Mo vibrational resonance. The population and relaxation of different dressed manifolds can produce unequal sideband intensities, diminish the symmetric partners, and broaden spectral profiles, particularly in a molecular system in an open space at room temperature that is subject to structural and relaxation inhomogeneity and molecular dynamics. Accordingly, the spectra are naturally deviated from the ideal symmetric Rabi doublet or standard Mollow triplet in a vibration-cavity coupling system. The central quantitative test is to establish the predicted nonlinear $\sqrt{n}$ scaling for the sideband displacements across different compounds, samples, and excitation conditions.

**Ladder of Mollow-type Raman sidebands in crystalline $Mo_2$ complexes.** The Raman spectrum of a single crystal of $Mo_2(DAniF)_4$ (**2**) provides a particularly clear view of the sideband structure associated with the Mo–Mo stretching vibration. As shown in Figures 4A and S4, the central Mo–Mo band at 406 cm$^{-1}$, corresponding closely to 405 cm$^{-1}$ for the powdered **2** (Figure 2B), is split into two peaks at 401 and 412 cm$^{-1}$, and is assigned to the bare vibrational resonance, $\omega_v$, in the single crystal environment. We tentatively attribute the 11 cm$^{-1}$ splitting of the $\omega_v$ band to ground state shifting caused by virtual excitation.[7] The $\omega_v$ position differs slightly from that observed for the powdered sample, consistent with differences in crystal packing, orientation, and local lattice environment. Despite the shift and splitting of the $\omega_v$, the spectrum retains a well-defined set of additional features distributed around the Mo–Mo resonance. The primary low-order structure consists of two components at 380 and 432 $cm^{-1}$, symmetrically displaced from the $\omega_v$ mode (406 $cm^{-1}$) by ± 26 $cm^{-1}$. They are therefore assigned as the Mollow-type sidebands, $\omega_\pm = \omega_v \pm \Omega_v$, with a fundamental branch displacement of $\Omega_v$ = 26 $cm^{-1}$. In addition, three blue sidebands appear at 475, 519 and 555 $cm^{-1}$, and a red sideband at 342 $cm^{-1}$. These blue Raman shifts ($\omega_{SB,n}$) correspond to effective manifolds of indices $n$ = 7, 19 and 33, respectively, while the red sideband matches the $n$ = 6 manifold (Figure 4A and Table S1). The consistency of the low- and high-order features with a single nonlinear coupling scale supports their organization into a ladder structure of dressed vibration–field states. For this complex, the featured Raman spectrum was reproduced recently by independent work from another group.[33] Under different experimental conditions, complex **2** exhibits a Mo–Mo stretching mode near 400 $cm^{-1}$ in the Raman spectrum, accompanied by bands at approximately 374 and 426 $cm^{-1}$.[33] This 400 $cm^{-1}$ mode is split into 406 and 395 $cm^{-1}$(Figure S6), consistent with our measurements of the single crystal sample. These features again define a primary splitting of about 26 $cm^{-1}$ on either side of the central resonance. The additional bands at 469, 513, and 547 $cm^{-1}$ correspond to larger displacements that can be assigned to higher coupling manifolds, with approximate indices $n$ = 7, 19, and 32 (Table S1), respectively. Surprisingly, although the central scattering is red shifted by 6 $cm^{-1}$ compared to our measurements, the fundamental coupling scale and the overall ladder structure remain nearly identical.

Similar Raman patterns are observed for microscrystalline **2** by changing the excitation condition from 532 nm laser of 25% to 405 nm laser of 10%. Specifically, the Raman spectrum recorded by excitation at 405 nm, which lies above the $\delta \rightarrow \delta^*$ electronic transition at approximately 450 nm (Figure 3),[24,27,3141] nearly reproduces the main sideband pattern observed with 532 nm excitation (Figures 4C and S5). In both spectra (Figures 4B and 4C), the principal Mo–Mo-centered components remain almost identical frequencies, and the higher-order sidebands are preserved. The 332 $cm^{-1}$ band for the 405-nm excited spectrum, which is weak in the 532-nm excited spectrum, is assigned to the vibro-polaritonic manifold of $n$ = 8 (Table S1). The persistence of the spectral pattern under increased excitation energy and reduced power indicates that the coupling signatures are not produced simply by local heating, photodegradation, or a nonlinear response that appears only at high incident power. Instead, these results indicate that incident excitation does not change the vibro-polaritonic manifolds but may modify the spectrum through population of the dressed states, thus, supporting the proposed Stokes Raman scattering coupling mechanism. The overall similarity of the spectra obtained under resonant and non-resonant conditions shows that the sideband positions are governed primarily by the Mo–Mo vibrational resonance and its resonant coupling to the local scattering field with $\omega_R \approx \omega_V$, rather than by the wavelength and power intensity of laser excitation.

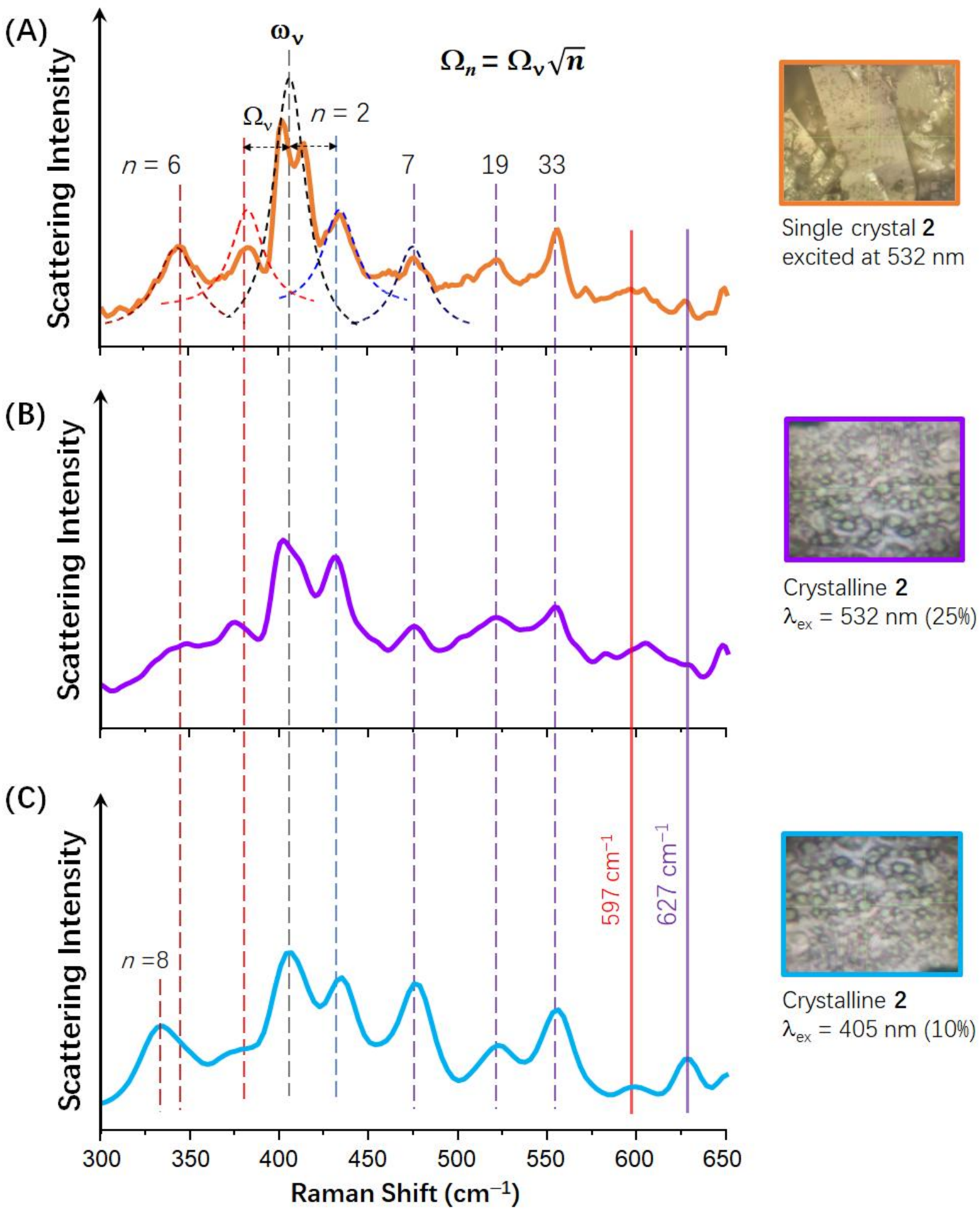

**Figure 4 . Multiplicity and reproducibility of Mollow-type Raman sidebands in**

**$Mo_2(DAniF)_4$ (2).** (A) Raman spectrum of a single crystal of **2** obtained with 532 nm excitation, showing the Mo–Mo-centered sideband ladder assigned according to $\Omega_n = \Omega_v\sqrt{n}$. (B) Raman spectrum of microcrystalline **2** obtained with 532 nm excitation at 25% laser power. (C) Raman spectrum of microcrystalline **2** obtained with 405 nm excitation at 10% laser power. The high-energy features at 597 and 627 $cm^{-1}$ are assigned to the leapfrog transitions within the dressed vibro-polaritonic ladder (see text).

Two additional high-energy Raman features at approximately 597 and 627 $cm^{-1}$ are reproducibly observed for complex **2** under different excitation conditions (Figure 4). These two bands lie beyond the principal sideband range associated with the lower-order dressed manifolds and can be alternatively interpreted in terms of leapfrog-type transitions arising from photon correlations within the vibro-polaritonic ladder.[36] In the dressed-state picture, such a pathway involves adjacent manifolds $n$ and $n+1$ connected through a virtual intermediate state, so that the observed Raman shift is determined by the total energy and the initial and final states, rather than by a sequential cascade through real intermediate states. For these two high-energy Raman shifts, the leapfrog transitions involve simultaneous emission of two photons from the red sidebands,[36] i.e., $|-\rangle \rightarrow |\pm\rangle \rightarrow |+\rangle$, and the energy correlation can be expressed by

$$\omega_{\mathrm{LF}} = (\omega_\nu - \Omega_n) + (\omega_\nu - \Omega_{n+1} + \Omega_n) = 2\omega_\nu - \Omega_{n+1},$$

with $\Omega_n = \Omega_\nu\sqrt{n}$ in the present ladder structure (Figure 3). Using $\omega_\nu = 406\ \mathrm{cm}^{-1}$ and $\Omega_\nu = 26\ \mathrm{cm}^{-1}$, the observed 627 $cm^{-1}$ feature is reproduced for $n = 50$, giving $\omega_{\mathrm{LF}} = 626.3\ \mathrm{cm}^{-1}$, whereas the 597 $cm^{-1}$ feature is reproduced for $n = 67$, giving $\omega_{\mathrm{LF}} = 597.6\ \mathrm{cm}^{-1}$. For both leapfrog processes, the calculation residuals are within the $\pm$ 1.0 $cm^{-1}$ assignment criterion used throughout this work. The persistence of these high-energy bands, together with their enhancement under higher-energy excitation, is therefore consistent with an additional relaxation pathway involving virtual intermediate states in the dressed vibration–field ladder. Although these peaks are also numerically matched by very high-order direct sideband assignments, population of non-resonant Raman transition to the single vibro-polaritonic manifolds is energetically inaccessible. A direct Mollow-sideband interpretation for the 597 $cm^{-1}$ and 627 cm$^{-1}$ bands would require uncoupled Raman-field energies near 465 and 347 nm, respectively. Such manifolds cannot be populated directly by non-resonant 532 nm Raman excitation, and the latter lies even above the exciton-polariton energy near 365 nm.[27] The leapfrog transition therefore accounts well for both high-energy scattering bands under non-resonant Raman excitation, i.e., at 532 nm. The agreement of these high-energy Raman shifts with the leapfrog-transition energy relation provides further support for the Mollow-type dressed-state structure and for the proposed coupling of the Mo–Mo vibration to the locally confined Stokes Raman field.

**Reanalysis of reported resonance-Raman shifts for $Mo_2(C{\equiv}CSiMe_3)_4(PMe_3)_4$.** A ladder-like pattern of Mo–Mo-centered Raman features has also been reported for another quadruply bonded $Mo_2(C{\equiv}CSiMe_3)_4(PMe_3)_4$.[34] In THF, the resonance-Raman spectrum exhibits two prominent bands at 362 and 397 cm$^{-1}$, together with a weak low-frequency band at 254 cm$^{-1}$, which were denoted $\nu_a$, $\nu_b$, and $\nu_c$, respectively. These bands were previously assigned to mixed skeletal normal modes containing contributions from $\nu(Mo_2)$, $\nu$(Mo–C), and Mo–C≡C bending coordinates, while significant mixing between the two dominant vibrational modes $\nu(Mo_2)$ and $\nu$(C≡C) was excluded. [34] Red shifts of the Raman features were observed upon substitution of the natural-abundance alkynyl group by the $^{13}C{\equiv}^{13}C$ isotopomer, which was considered to be evidence for participation of the Mo–C≡C skeleton. However, the calculated mixed-mode frequencies and potential-energy distributions derived from the valence force-field normal-coordinate model are not uniquely determined and do not provide a strict quantitative reproduction of the observed Raman shifts. More importantly, this assignment does not identify an independently observed Mo–Mo stretching fundamental, even though the Mo–Mo stretch is expected to be one of the most characteristic low-frequency Raman-active modes of quadruply bonded $Mo_2$ complexes.[31] We therefore reexamine these reported spectra within the framework of Stokes Raman scattering coupling.

In this alternative interpretation, the two intense bands at 362 and 397 cm$^{-1}$ are assigned to the lower and upper vibro-polaritonic branches, $P_\nu^-$ and $P_\nu^+$, of an effective Mo–Mo-centered vibrational mode (Table S2). The center frequency is then approximately 380 cm$^{-1}$, and the $P_v^+ - P_v^-$ separation gives the vacuum Rabi splitting of 35 cm$^{-1}$, comparable to the coupling strengths ($\Omega_v$) observed for the $Mo_2$ formamidinate complexes. The weak band at 254 cm$^{-1}$ can be assigned to a red sideband predicted by $\Omega_n\sqrt{n}$ with $n$ = 13, while the band at 619 cm$^{-1}$, previously assigned as a mixed combination feature, corresponds to a high-order blue sideband with $n$ = 47. The $^{13}C{\equiv}^{13}C$ isotopomer shows the same behavior after isotope-induced shifting of the effective Mo–Mo-centered coordinate. For this $^{13}C{\equiv}^{13}C$ containing complex, the spectrum exhibits two intense bands at 355 and 387 cm$^{-1}$, giving an effective center frequency of 371 cm$^{-1}$ corresponding to the Mo–Mo stretching mode and a reduced coupling scale of about 32 cm$^{-1}$. The reported bands at 247 and 599 cm$^{-1}$ can then be assigned to the corresponding red and blue sidebands with $n$ = 15 and 51 (Table S2), respectively. Thus, isotope substitution shifts the effective vibrational coordinate and coupling strength, while the sideband positions shift consistently with the Stokes Raman scattering coupling model. The intense Raman scattering and resolved coupling behavior of this alkynyl $Mo_2$ complex may further reflect the increased polarizability of the $Mo_2$ core induced by the electron-rich, π-conjugated alkynyl ligands, consistent with the ligand-dependent polarizability effect discussed above for the formamidinate complexes.

A key issue with the mixed-normal-mode assignment is the apparent absence of an isolated Mo–Mo stretching fundamental. In the present interpretation, this apparent absence of the $\omega_v$ mode can be understood from resonant coupling of a bare molecular vibration under resonance Raman excitation, which alters both the electronic and vibrational structures.[8] Consequently, the Mo–Mo mode is redistributed into lower and upper vibro-polaritonic branches and, under stronger local-field conditions, into Mollow-type sidebands. The isotopic effect naturally shifts the effective Mo–Mo-centered coordinate, but does not necessarily influence each independent static mixed normal mode. In contrast, the Stokes Raman scattering coupling model simultaneously accounts, with quantitative agreement (Table S2), for the intense doublet, the lower and upper sidebands in the proximity of the $\nu(Mo_2)$, the isotope-dependent shift of the effective center frequency, and the absence of an isolated bare Mo–Mo fundamental. The high-*n* sidebands require an additional energy-access pathway because they would not be accessible energetically if the Raman field were generated only from the bare $\delta \rightarrow \delta^*$ resonance at 648.7 nm.[34] In a previous study, we have demonstrated that resonance excitation of the $\delta \rightarrow \delta^*$ transition in complex **2** induces coupling to the locally confined Rayleigh/scattering field of the $Mo_2$ core, forming the exciton-polariton states.[27] A plausible rationalization for the appearance of the high-energy sidebands is that the upper exciton-polariton resulting from resonant Raman excitation may provide a higher-energy hybrid scattering channel, allowing access to the high vibration-field coupling manifolds. This explanation is consistent with the formation of polaritonic potential-energy surfaces involving electronic, nuclear, and photonic degrees of freedom.[8,9] Therefore, the reported Raman spectra of $Mo_2(C{\equiv}CSiMe_3)_4(PMe_3)_4$ are more naturally explained from Raman scattering from dressed Mo–Mo vibrational states than as a set of unrelated mixed skeletal fundamentals.

**Confinement and enhancement of the Raman scattering field by the $Mo_2$ optical resonator.** In inelastic Raman scattering, interaction of a molecule with the incident laser photons generates a weak Stokes field whose frequency shift reports the molecular vibration energy. For quadruply bonded $Mo_2$ complexes, a single Raman band near 400 $cm^{-1}$ may therefore be expected, representing the Mo–Mo stretching mode ($\omega_v$), as observed for $Mo_2(O_2CCH_3)_4$ (**1**). By contrast, this study reveals a Rabi-type doublet and Mollow-type sidebands centered at $\omega_v$ in the Raman spectra for the dimolybdenum formamidinates **2** and **2′** and $Mo_2(C{\equiv}CSiMe_3)_4(PMe_3)_4$, which commonly possess a polarizable $Mo_2$ unit supported by electron donating and/or resonant ligands. Observation of these features require coherent mixing between the vibrational excitation and an intense EM field, as observed in the vibration-cavity coupling. Under the Raman spectroscopic condition without an external optical cavity, the Stokes scattering induced by two-phton Raman transition[18,22,23] is the only EM source possibly involving in the molecule-photon interaction. However, this scattering field must be sufficiently enhanced for coherently coupling the vibrational normal mode, which is potentially enabled by the confinement of the local scattering by the $Mo_2$ unit that acts as a diatomic optical resonator. This potential of a $M_2$ unit in general has been demonstrated in exciton-scattering field coupling in our previous studies.[24,25,26,27] The observed vibration-field interaction here is consistent with coherent coupling of a two-level electronic excitation and the $M_2$ enhanced scattering field in the visible region. The phenomenon of enhancing the Raman scattering field by the $Mo_2$ is in analogy to tip- and surface- enhanced Raman scattering, specifically in systems at the angstrom scale: strong Raman enhancement originates from highly localized EM fields produced by plasmonic nanostructures, metallic junctions, or sharp metallic tips.[18,19,23,42] Given the angstrom set up of two metal atoms fixed by the ligand framework, this $M_2$ resonator advantageously operates under ambient, cavity-free Raman conditions. Therefore, observation of the vibro-polaritonic Raman signals provides a strong evidence supporting the functionality of a ligand-supported $M_2$ unit as an diatomic resonator, profoundly different from picocavities with "naked" metallic atoms on the tip.[12,13,19]

This interpretation is strongly supported by the photon-number-dependent sideband structure and the observed coupling strengths following the nonlinear scaling $\Omega_n = \Omega_v\sqrt{n}$ with the $n$ index corresponding to the photon number state.[37,38,39] The appearance of higher-order blue sidebands shows that Raman transitions can populate the dressed vibrational states beyond the lowest vibro-polaritonic doublet or triplet. Such behavior is difficult to reconcile with ordinary spontaneous Raman scattering from independent molecules, but is consistent the nature of the hybrid molecular system with a strongly enhanced and spatially confined Raman field generated within the $Mo_2$ molecular framework.[3,9,18] This molecular-resonator picture extends the concept of atom-scale field confinement from plasmonic picocavities to a chemically defined metal–metal bonded $Mo_2$ unit, approaching the structural limit of optical confinement in a real molecular system.

## Discussion

Raman spectroscopy of quadruply bonded $Mo_2$ complexes reveals that the Mo–Mo stretching vibration ($\omega_v$) around 400 cm$^{-1}$ can be strongly and resonantly coupled to the locally generated Stokes Raman scattering field that is generated, confined, and enhanced by the $Mo_2$ unit. In conventional Raman scattering, an energy ($n\hbar\omega_v$) is transferred from the incident optical field to the molecular vibration and is reported as the corresponding Stokes Raman shift. In the present $Mo_2$ system, we propose that this Raman energy-transfer channel is converted into an effective local scattering field that is confined and enhanced by the dimetal unit. The fundamental Raman-field mode, ($\omega_R \approx \omega_v$), therefore becomes resonant with the Mo–Mo stretching transition, replacing the bare vibrational states with dressed vibro-polaritonic manifolds. Consistent with this picture, the more polarizable formamidinate complexes $Mo_2(DAniF)_4$ and $Mo_2(DTolF)_4$ exhibit Rabi-type splitting, Mollow-type sidebands, and higher-order Raman features, whereas $Mo_2(O_2CCH_3)_4$ shows essentially a single Mo–Mo stretching band.

The quantitative organization of the sideband displacements according to $\Omega_n = \Omega_v\sqrt{n}$ provides the central evidence for a Jaynes–Cummings-type ladder of dressed vibration–field states, in which the Raman/vibrational quantum index ($n$) maps onto the photon occupation of the effective local Raman mode. The reproducibility of the sideband positions across different solid-state samples and excitation conditions further indicates that the ladder is governed primarily by the Mo–Mo vibrational resonance and its coupling to the local scattering field rather than by the absolute excitation wavelength. The unusually high-energy Raman features at 597 and 627 cm$^{-1}$ require a distinct relaxation pathway: their direct assignment as ordinary high-($n$) Mollow sidebands would imply energetically inaccessible manifolds under nonresonant excitation, whereas both shifts satisfy the energy relation expected for leapfrog-type transitions involving adjacent dressed manifolds and a virtual intermediate state. Under resonant excitation, the upper electronic exciton-polaritonic state may further facilitate access to high-order vibration–field manifolds. This Stokes Raman scattering coupling framework is supported by Reanalysis of previously reported resonance-Raman spectra of $Mo_2(C{\equiv}CSiMe_3)_4(PMe_3)_4$.

The present vibration–field coupling is very similar to the excitation–field coupling that was previously observed in the resonance fluorescence of dimetal molecular resonators. Together, these electronic and vibrational manifestations support a unified molecular resonator model, in which the $Mo_2$ unit acts simultaneously as a two-level emitter and a Raman-active oscillator, meanwhile, being the source of the local scattering field, which is extremely confined and enhanced by the ångström-scale structure. More broadly, this coupling model sheds new light on the photoluminescence and Raman spectroscopy of metal–metal bond chemistry, where split bands, satellite features, and high-order spectroscopic features which have been interpreted in different ways from a chemical perspective. Treating the metal–metal vibration as a molecular resonator and quantum oscillator coupled to its locally generated scattering field introduces an additional spectroscopic dimension. Therefore, the results therefore extend optical-field confinement from externally fabricated nanocavities and picocavities to a chemically defined metal–metal bond, establishing quadruply bonded $Mo_2$ complexes as a molecular platform for cavity-free light–matter coupling under ambient conditions.

**Acknowledgments**

We acknowledge the primary financial support from the National Natural Science Foundation of China (22171107, 21971088, 21371074), Natural Science Foundation of Guangdong Province (2018A030313894), Jinan University, and the Fundamental Research Funds for the Central Universities.

**Author Contribution**

C.Y.L. conceived this project and designed the experiments and worked on the manuscript. M.M. carried out the major experimental work, and prepared the Supplementary Information. Y.N.T. and G.Y.Z. involved in the spectroscopic data analysis and assisted in manuscript preparation.

Competing interests: The authors declare no conflict of interest.

[38] Brune, M., Schmidt-Kaler, F., Maali, A., Dreyer, J., Hagley, E., Raimond, J. M., and Haroche, S. Quantum Rabi oscillation: A direct test of field quantization in a cavity. *Phys. Rev. Lett.* **76,** 1800–1803(1996).

[39] Fink, J. M., Goeppl, M., Baur, M., Bianchetti, R., Leek, P. J., Blais, A., Wallraff, A. Climbing the Jaynes–Cummings ladder and observing its $\sqrt{n}$ nonlinearity in a cavity QED system. *Nature*, **454**, 315–318 (2008).

[40] Sideband assignments in this and the related $Mo_2$ systems were made by selecting the integer photon-manifold index n that minimizes the difference between the observed Raman position and that calculated from the coupling relation $\Omega_n = \Omega_v\sqrt{n}$. An assignment was retained only when $|\omega_{SB,n} - [\omega_v \pm \Omega_n(cal)]| \leqslant 1.0\ cm^{-1}$. The observed and calculated sideband positions and their residuals are listed in Tables S1 and S2.

[41] Lin, C., Protasiewicz, J. D., Smith, E. T. & Ren, T. Linear free energy relationships in dinuclear compounds. 2. Inductive redox tuning via remote substituents in quadruply bonded dimolybdenum compounds. Inorg. Chem., 35, 6422–6428 (1996).

[42] Verma, P. Tip-Enhanced Raman Spectroscopy: Technique and Recent Advances. Chem. Rev. **117**, 6447–6466 (2017).